\documentclass[preprint]{JHEP3}
\usepackage{epsfig}
\usepackage{graphicx,psfrag}
\usepackage{dcolumn}
\usepackage{bm}
\usepackage{amsmath,amssymb}

\newcommand{\beq}{\begin{equation}}
\newcommand{\eeq}{\end{equation}}
\newcommand{\bea}{\begin{eqnarray}}
\newcommand{\eea}{\end{eqnarray}}
\newcommand{\hf} {\frac{1}{2}}
\newcommand{\nonu}{\nonumber\\}
\newcommand{\nn}{\nonumber\\}

\newcommand{\sla}[1]{\mbox{$#1\!\!\!/$}}

\newcommand\eqn[1]     {Eq.\,(\ref{#1})}
\newcommand\eqns[2]    {Eqs.\,(\ref{#1}) and~(\ref{#2})}

\newcommand\fig[1]     {Fig.\,{\ref{#1}}}
\newcommand\sect[1]    {Sect.\,{\ref{#1}}}

\def\ord#1{{\cal O}(#1)}

\def\mr#1{{\mathrm{#1}}}
\def\ci{{\rm i}}

\def\t{\tilde}
\def\eq#1{(\ref{#1})}

\title{Renormalization of QCD$_2$}

\author{J. Kov\'acs$^a$, S. Nagy$^a$, I. N\'andori$^b$, K. Sailer$^a$\\
$^a$Department of Theoretical Physics, University of Debrecen,\\ P.O. Box 5, H-4010 Debrecen, Hungary\\
$^b$Institute of Nuclear Research,\\ P.O.Box 51, H-4001 Debrecen, Hungary}
\abstract{
The low energy infrared scaling of the multi-color 2-dimensional quantum
chromodynamics is determined in the framework of its bosonized model
by using the functional renormalization group method in the local potential approximation.
The model exhibits a single phase with a superuniversal effective potential.
}
\keywords{Renormalization group}

\preprint{}

\begin{document}

\section{Introduction}

The low dimensional fermionic models as toy models provide an excellent playground to
try and develop new ideas and methods in quantum field theory \cite{Abdalla}.
These models have only indirect physical meaning but they are much simpler than
their 4-dimensional counterparts, and they usually show important characteristics of the original ones.
For example the 2-dimensional quantum electrodynamics (QED$_2$) proved to be a good toy model
for treating the soft mechanism of the quark confinement \cite{Fischler,Nagy_QED2}.
The confining properties, the large-N$_c$ expansion \cite{THooft} or the baryon structure \cite{Frishman,Blas}
of QCD can also be studied in the 2-dimensional version of the model and then one
can get even analytical results for the non-perturbative domain.
One usually takes the bosonized version of these models which are local self-interacting
scalar theories, and can be investigated in an easier way \cite{Coleman}.

The phase structure of the QED$_2$ with many flavors was mapped out from its bosonized version
and it was shown that it exhibits only a single phase \cite{Nagy_MQED2,Nandori_Mschw} as opposed to the
single-flavor QED$_2$ (which is often referred to as the massive Schwinger model)
\cite{Coleman,Nagy_QED2}, which possesses a symmetric strong coupling
($e \gg m_e$) phase and the weak coupling ($e \ll m_e$) phase with spontaneously broken
reflection symmetry separated by the critical value $(m_e/e)_c\sim 0.31$ as was
shown by density matrix renormalization group (RG) technique \cite{dm_rg} or by continuous
RG method \cite{Nagy_MSG,Nandori_Mschw}.

The situation is a bit different in the case of the 2-dimensional quantum chromodynamics (QCD$_2$).
The scalar model equivalent of the single flavor QCD$_2$ can be easily obtained,
but the bosonization failed to treat the multi-flavor model.
The solution of this problem was the introduction of the non-abelian bosonization \cite{Witten}.
However a single-flavor but multi-color investigation is possible with the usual
abelian bosonization technique.

It is argued \cite{Baluni} that the multi-color QCD$_2$ possesses two phases, a weak coupling or quark phase,
and a strong coupling or Bose phase. Elsewhere it is also argued \cite{singlephase} that the model has a single
phase. This open question should be investigated in the low energy limit.

The bosonized version of the multi-flavor QED$_2$ contains sine-Gordon (SG) type
periodic self-interaction terms. The scalar fields are coupled by a mass matrix giving
a multi-component  or layered sine-Gordon (LSG) model
which is used to describe the vortex dynamics of magnetically coupled
layered superconductors \cite{LSG}, where the number of flavors in QED$_2$ equals the number of
layers of the condensed matter system \cite{Nandori_Mschw}.
The Bose form of the multi-color QCD$_2$ also contains SG type interactions, a mass
matrix and a mixed term. The latter can be associated to the non-periodic term in the potential
therefore it can be Taylor expanded giving further contributions to the mass matrix.
The higher order terms are negligible since they do not modify the phase structure of the model.
Then one can conclude that the difference between the bosonized versions of
QED$_2$ and QCD$_2$, respectively stems from the different mass matrices.
For the  2-flavors ($N_f=2$) QED$_2$ and the 2-colors ($N_c=2$) QCD$_2$ the mass 
matrices coincide, implying that these models are equivalent. It is quite surprising
since the fermionic models are different.
The difference between the bosonized models appears when $N_f>2$ and $N_c>2$.
We note that the low energy multi-flavor QCD$_2$ with unequal masses
can also be bosonized giving a so called generalized SG model
\cite{Blas,Blas2} but its investigation is out of the scope of this paper.

The phase transition of these models was obtained from the microscopic theory so far,
which is formulated in the high energy/ultraviolet (UV) region.
The low energy/infrared (IR) physics can be obtained by integrating out the
quantum fluctuations one by one. Then one can get a low energy theory describing
the quantum system at energy scales where the measurements are usually performed.
The quantum fluctuations can be eliminated systematically by using the renormalization
group (RG) method. The original fermionic models contain strong couplings in the high energy UV regime which
disables one to perform a perturbative renormalization. The evolution is usually
started from a perturbative region where the theory is almost interaction free.
We also note that the RG equations should preserve the gauge symmetry \cite{gauge-RG}.
However the bosonized version of the toy models, which are simple scalar models
can be easily treated by the functional RG method.
The low energy IR physics of the 2-dimensional one-component scalar field theories
which contain periodic self-interaction term are well understood
\cite{Nandori_Mschw,Nagy_MSG,Nagy_SG,Nagy_ZSG}.

Our goal in this article is to compare the phase structure of the multi-flavor QED$_2$
and the multi-color QCD$_2$ by using functional RG method.
The bosonization is applicable for the original fermionic models at a certain parameter
choice $\beta^2=4\pi$, which appears in the argument of the sine function.
Therefore the value of the parameter $\beta$ has to be kept fixed during the RG evolution.
It implies that the investigation which is confined to the local potential approximation (LPA)
where $\beta$ does not evolve may give reliable evolutions. Although the Wilsonian
renormalization procedure is very powerful in LPA \cite{Hasenfratz}, we choose the effective average action
RG method \cite{Berges,effac,CS,Nagy_ZSG} to obtain the evolution of the models due to its
more flexible usage. The RG evolution provides us the scale dependence of the couplings in the scalar model from
which the scaling of the original fermionic couplings can be obtained according to the bosonization
rules.

The paper is organized as follows. In \sect{sec:boso}  we introduce the bosonized versions
of the investigated fermionic models and relate them to  layered sine-Gordon (LSG) models.
The evolution of the couplings is determined in \sect{sec:ren}.
Finally, in \sect{sec:sum} the conclusions are drawn up.

\section{Bosonized models}\label{sec:boso}

\subsection{Multi-flavor QED$_2$}\label{sec:QED2}

The Lagrangian of the multi-flavor QED$_2$ is
\beq
{\cal L}=-\frac14 F_{\mu\nu}F^{\mu\nu}+\sum_{i=1}^{N_f} \bar\psi_i\gamma^\mu(\partial_\mu-
\ci e A_\mu)\psi_i-m_e \sum_{i=1}^{N_f} \bar\psi_i\psi_i,
\eeq
with $N_f$ Dirac fields and identical fermionic charge $e$ and mass $m_e$, furthermore
$F_{01} = \partial_0 A_1-\partial_1 A_0$. One can transform the fermionic field variables
$\bar\psi_i,\psi_i$ into bosonic ones $\phi_j$ by the bosonization rules
\cite{boson,Coleman 1976,Fischler}
\bea
:\bar\psi_i\psi_i: &\to& -\frac{cm_e e}{\sqrt\pi}\cos(2\sqrt{\pi}\phi_i), \nonu
:\bar\psi_i\gamma_5\psi_i: &\to& -\frac{cm_e e}{\sqrt\pi}\sin(2\sqrt{\pi}\phi_i), \nonu
:\bar\psi_i\gamma_\mu\psi_i: &\to&
 \frac1{\sqrt{\pi}}\varepsilon_{\mu\nu}\partial^\nu \phi_i, \nonu
:\bar\psi_i\ci \sla\partial\psi_i: &\to& \frac12 {\cal N}_{m_e} (\partial_\mu\phi_i)^2,
\eea
where ${\cal N}_{m_e}$ means normal ordering with respect to the fermion mass $m_e$ and $c=\exp(\gamma)/2\pi$,
with the Euler constant $\gamma=0.5774$. The Hamiltonian of the system in Coulomb gauge is given by
\beq
{\cal H} = \sum_{i=1}^{N_f} \int_x \bar\psi_i(x)(\ci \gamma_1\partial_1+m_e)\psi_i(x)
-\frac{e^2}{4}\int_{x,y} j_{0,x}|x-y|j_{0,y},
\label{hamcoul}
\eeq
with $\int_x=\int_0^T dx^0 \int_{-L}^L dx^1$ and
\beq
j_{0,x} = :\sum_{i=1}^{N_f} \bar\psi_i(x)\gamma_0\psi_i(x): =\frac1{\sqrt{\pi}}\partial_1\sum_{i=1}^{N_f} \phi_i(x).
\eeq
The resulting bosonized form of the Hamiltonian is
\bea
{\cal H} &=&
{\cal N}_{m_e} \int_x \biggl[\frac12\sum_{i=1}^{N_f} \Pi^2_i(x)+\frac12\sum_{i=1}^{N_f}(\partial_1\phi_i(x))^2
+\frac{e^2}{2\pi}\left(\sum_{i=1}^{N_f} \phi_i(x)\right)^2\nn
&&~~~~~~~~~~~~-cm_e^2\sum_{i=1}^{N_f} \cos\left(2\sqrt\pi\phi_i(x)\right)\biggr],
\label{hamsc}
\eea
where $\Pi_i(x)$ denotes the momentum variable canonically conjugated to 
$\phi_i(x)$.  In order to complete the bosonization, one has to use normal-ordering 
with respect to the scalar mass $\mu^2=e^2/\pi$ which modifies the coupling of the 
periodic term, ${\cal N}_{m_e} \cos(b\phi) = (\mu/m_{e})^{b^2/4\pi} {\cal N}_{\mu} \cos(b\phi)$.
Therefore, the $N_f =1$ flavor bosonized QED$_2$ reads as
\beq
{\cal H}_{N_f =1} = {\cal N}_{\mu} \int_x \left[\frac12  \Pi^2(x)+\frac12 (\partial_1\phi(x))^2
+\frac{1}{2} \mu^2  \phi(x)^2
-c m_e \mu  \cos\left(2\sqrt\pi\phi(x)\right)\right],
\eeq
which can be generalized for $N_f >1$ flavor using its rotated form where the 
mass matrix is diagonal. Let us note that the scalar mass term in \eq{hamsc} can be 
rewritten in terms of the mass matrix ${\cal{M}}_{\mr{QED}}^2$ defined via
\begin{equation}\label{mqed1}
\hf \Phi {{\cal M}^2_{\mr{QED}}} \Phi 
= \frac{1}{2} \frac{e^2}{\pi} \left(\sum_{n=1}^{N_f} a_n \phi_{n}\right)^2,
\end{equation}
where $\Phi = (\phi_1, \phi_2, ..., \phi_{N_f})$ and the couplings $a_n$ are free real
parameters of the model.  In order to reproduce the mass term of \eq{hamsc} 
one has to restrict the choice to be $a_n =1$ for all $n$. However, based on 
symmetry considerations any choice with $a^2_n = 1$ for all $n=1,\ldots,N_f$ 
should reproduce exactly the same phase structure since the number of 
zero and non-zero eigenvalues of the mass-matrix remains unchanged
which was found to be decisive with respect to the phase structure of the 
$N_f$-component model \cite{Nandori_Mschw}. It is not a surprise since in path-integral
quantization a change of sign of any of the field components represents an allowed
transformation of the integration variable, that in turn induces a single change of
sign in the mass term and leaves the other terms of the action invariant.
As a consequence, $a_n = (-1)^{n+1}$ is also a suitable choice which will be used in the
mass term of QED$_2$ in this paper and leads to the mass matrix
\beq\label{mass_qed}
({\cal M}_{\rm{QED}}^2 )_{a,b}= (-1)^{a+b} G, ~~~~a,b=1,2,\ldots, N_f
\eeq
with $G= e^2/\pi$. The QED-type mass matrix \eq{mass_qed} exhibits
a single non-vanishing mass-eigenvalue, $M^2_{N_f} = N_f G$ and $N_f-1$ vanishing eigenvalues.

\subsection{Multi-color QCD$_2$}\label{sec:QCD2}

The Hamiltonian of the QCD$_2$ with a single flavor $N_f=1$ is
\beq
{\cal H} = g^2\sum_{a,b=1}^{N_c}E_a^{b~2}+\sum_{a,b=1}^{N_c}
\bar\psi^{a}\gamma_1(i\delta_a^b\partial_1-A_a^b)\psi_{b}
+m_g \sum_{a=1}^{N_c}\bar\psi^{a}\psi_{a}
\eeq
in the gauge
\bea
A_0 = 0,~~
A_b^a = 0\mbox{~for~} a=b,~~
E_b^a = 0\mbox{~for~} a\ne b.
\eea
Using the Gauss law the bosonized Hamiltonian with one flavor becomes
\bea\label{qcdbos}
{\cal H} &=& \sum_a\left[\hf\left(\Pi_a^2+(\partial_1 \phi_a)^2\right)
-\frac{c m_g \mu}{\pi}{\cal N}_{\mu}\cos(2\sqrt{\pi}\phi_a)\right]\nn
&&+\frac{g^2}{8\pi N_c}\sum_{a,b}(\phi_a-\phi_b)^2+\frac{2c^2\mu^2}{\pi^{3/2}}\sum_{a,b}
\frac{\sin(2\sqrt{\pi}(\phi_a-\phi_b))}{\phi_a-\phi_b},
\eea
where the scale $\mu$ should satisfy $\mu = c' g$, with $c'$ a constant, in order to take the
interaction energy proportional to $g^2$ \cite{Baluni}.

We determined the field configuration for the ground state of the model numerically. The  static
field configuration  minimizing  the energy was searched for by means of the conjugate gradient method 
as in \cite{Nagy_QED2}. The results showed that the constant field configuration with all components
$\phi_a$ equal to the same constant minimizes the energy, for the
cases $N_c=2,3$. The same trivial ground state appears when we take the polynomial
piece of the potential alone. Therefore, it seems to be justified to Taylor-expand
the last potential term in the Hamiltonian at $\phi_a-\phi_b=0$ and keep only 
the quadratic term of the expansion, so far one is interested in the behavior of the system in
or close to the ground state. Then one finds
\bea\label{qcdbosexp}
{\cal H} &=& \sum_a\left[\hf\left(\Pi_a^2+(\partial_1 \phi_a)^2\right)
-\frac{c c' m_g g}{\pi}{\cal N}_\mu\cos(2\sqrt{\pi}\phi_a)
+ \hf \sum_b g^2 c_g(\phi_a-\phi_b)^2\right]
\eea
with $c_g=1/4\pi N_c+(4cc')^2/3$. Let us note, that similarly to the bosonized QED$_2$, the
scalar mass term in \eq{qcdbosexp} can be rewritten in terms of the mass matrix
${\cal{M}}_{\mr{QCD}}^2$ defined via 
\bea
\label{mqcd1}
\hf \Phi {\cal M}_{\rm{QCD}}^2 \Phi   
=\hf g^2c_g \sum_{a,b}   (\phi_{a} -\phi_{b})^2,
\eea
where $\Phi = (\phi_1, \phi_2, ..., \phi_{N_c})$ and the summation runs 
from $a,b=1$ to $N_c$. Then one gets
\beq
\label{mass_qcd}
({{\cal M}}^2_{\mr{QCD}})_{a,b}= (N-1)J\delta_{a,b} - J~~~~a,b=1,2,\ldots, N_c
\eeq
 with $J=2g^2c_g $. The QCD-type mass matrix \eq{mass_qcd} has a single zero 
eigenvalue and $N_c-1$ identical, non-vanishing eigenvalues, $M^2_{N_c}= N_cJ$.

\subsection{Relation to layered sine-Gordon models}\label{sec:LSG}

Both models, the bosonized multi-flavor QED$_2$ and the multi-color QCD$_2$ can 
be considered as the specific forms of a generalized LSG model \cite{Nandori_Mschw}
which consists of two-dimensional periodic scalar fields coupled by an appropriate mass 
matrix whose bare Euclidean action is written as
\beq
\label{lsg}
S = \int_x \left[ \hf (\partial_\mu \Phi)^2
+\hf \Phi {\cal M}^2 \Phi + y \sum_{n=1}^N  \cos (\beta \phi_n)
\right]
\eeq
with the $O(N)$ multiplet $\Phi=\left(\phi_1, \dots, \phi_N\right)$. 
For the specific choice $\beta^2=4\pi$  with the mass matrices  \eq{mass_qed} and
\eq{mass_qcd} one one recovers the bosonized version of the multi-flavor QED$_2$ for
$N=N_f$ and that of the multi-color QCD$_2$ for $N=N_c$, respectively.
The amplitude $y$ of the periodic piece of the potential is  identical for all component fields,
 and it is proportional to the fermion mass  ($y \sim m$),
the exact  relation can be determined by using normal-ordering w.r.t. the 
boson mass. We note that for $N=2$ the mass matrices \eq{mass_qed} and 
\eq{mass_qcd} coincide, consequently, the Bose forms of the
two-flavor QED$_2$ and the two-color QCD$_2$ are the same.

For later use it is worthwhile mentioning that after an appropriate $O(N)$ rotation diagonalizing
the mass matrix, the LSG model with the QED-type mass matrix \eq{mass_qed} exhibits
a single massive field and $N-1$ massless ones. On the contrary, the  LSG model
with the QCD-type mass matrix  \eq{mass_qcd} shows up a single massless field  
and $N-1$ massive fields of identical masses after such an $O(N)$ rotation.

\section{RG approach for multi-component models}\label{sec:ren}

The systematic removal of the quantum fluctuations can be performed by
the evolution equation for the effective action \cite{Berges,effac,CS,Nagy_ZSG}
\beq\label{Wett}
k\partial_k \Gamma_k = \hf\mr{Tr}\frac{k\partial_k R_k}{\Gamma^{(2)}_k+R_k},
\eeq
where $\Gamma^{(2)}_k$ refers to the second functional derivative matrix of the effective action and
the trace Tr stands for the integration over all momenta. The scale $k$ starts from a large UV value
$\Lambda$ (which is typically set to $\infty$ during the calculations) and goes
to zero. $R_k$ plays the role of the IR regulator function. For the suppression of the high-frequency
modes one can choose the power-law type regulator
\beq
R_k = p^2\left(\frac{k^2}{p^2}\right)^b
\eeq
with the parameter $b\ge 1$. Here we choose $b=1$, which corresponds to the Callan-Symanzik RG scheme \cite{CS}.
It is easy to see, that in $d=2$ the chosen CS scheme is free of UV divergences and ultralocal,
furthermore the evolution equations take a rather  simple form. We note
that in $d=2$ the choices $b=1$ and $b=\infty$ coincide \cite{Nandori_scheme,Nandori_multiSG} in the LPA.
The latter case corresponds to the sharp cutoff limit, which makes the functional form
of the CS and the sharp cutoff (or Wegner-Houghton type) evolution equations similar.
The effective action is expanded in powers of the derivative of the field,
\beq
\Gamma_k[\Phi] = \int_x\left[V_k[\Phi]+Z_k[\Phi](\partial_\mu\Phi)^2+\ord{\partial_\mu^4}\right],
\eeq
with $V_k[\Phi]$ the potential  and $Z_k[\Phi]$ the wave-function renormalization.
The latter provides evolution to the parameter $\beta$, even in the case of
field-independent wave-function renormalization $Z_k[\Phi]\equiv z_k$, where $\beta^2=1/z$ \cite{Nagy_ZSG}.
The bosonization gives a constraint to the parameter $\beta^2=4\pi$. The running of $Z_k[\Phi]$
influences the evolution significantly in the vicinity of the Coleman point at $\beta^2=8\pi$
\cite{Nagy_ZSG,Nagy_2010} but gives slight modifications around $\beta^2=4\pi$
\cite{Nagy_MSG}, therefore it is not supposed to affect the phase structure of the model.
Thus we do not go beyond the LPA, and set $Z_k[\Phi]=1$. One can derive the evolution equation
\beq
\label{eveq}
(2+k \, \partial_k) \,\, \tilde V_k ({\Phi}) = 
- \frac{1}{4\pi} \ln \left[ {\mr{det}} \left(\delta_{ij} 
+ \tilde V_k^{ij}({\Phi}) \right) \right],
\eeq
with $  {\tilde V}_k^{ij}= \partial_{\phi_i}\partial_{\phi_j} {\tilde V}_k$
for the dimensionless potential ${\tilde V_k} = k^{-2} V_k$, where $\Phi$ stands for
homogeneous field configurations. We make the general ansatz
\begin{equation}
\label{def1}
{\tilde V}_{k}({ \Phi}) = 
\frac{1}{2} { \Phi} 
{ { {\tilde {\cal M}}}}^{2}(k) 
{ \Phi}
+{\tilde y}(k) \sum_{n=1}^N  \, \cos(\beta \, \phi_n)
\end{equation}
for the dimensionless potential of the LSG type models under discussion,
where $\tilde y(k) = k^{-2} y(k)$. Inserting the ansatz  \eq{def1} 
into \eqn{eveq}, the right-hand side becomes periodic, while 
the left-hand side contains both periodic and non-periodic parts 
\cite{Nandori_LSG,Nandori_JPA,Jentschura,Nandori_Mschw}. 
The non-periodic part contains only mass terms, so that we obtain a 
RG flow equation for the dimensionless mass matrix 
\begin{equation}
\label{mass_rg}
\left(2 + k\partial_k \right){ { {\tilde {\cal M}}}}^{2}(k) = 0, 
\end{equation}
giving the scaling
\beq
\tilde J_k = k^{-2} J, ~~\mbox{and}~~\tilde G_k = k^{-2} G,
\eeq
which corresponds to the scaling according to the canonical dimensions, since
in the LPA the anomalous dimension is zero. One can conclude that the dimensionful couplings 
$J$, $G$ remain constant during the blocking.

The RG flow avoids the singularities if we handle the evolution without any truncations \cite{Berges}.
We note however that one should usually use some approximations or expansions in the RG equations
in order to solve them. The bosonized QCD$_2$ contains a single Fourier mode, however the RG equations
generate the higher harmonics. Restricting ourselves to follow the evolution of the fundamental
mode only may induce a strong truncation,
implying that we should face the problem of poles of the evolution equation in \eqn{eveq}.
The proper choice of the IR regulator function may drive the evolution to reach the pole
only in the $k\to 0$ limit. This  seems to be true even in the case of the sharp cutoff
scheme, where one can draw up the quantum cenzorship conjecture \cite{Pangon_phi4,Pangon_SG}.
However the poles -- unless they appear as artifacts of the
approximations and truncations, --  have a great physical importance, because
their existence can signal the spontaneously broken phase of the model \cite{Berges,Litim}.
In the symmetry broken phase of a single component scalar field the 
dynamical Maxwell cut makes the effective potential superuniversal, namely
\beq\label{parab}
\tilde V_{k=0}[\Phi] = -\hf \Phi^2.
\eeq

\subsection{UV scaling}\label{uvscal}

The correct UV scaling can be obtained if we improve the results of the linearized approximation
by taking into account corrections of the order ${\cal O}( \t J)$ 
for the QCD$_2$ type and those of ${\cal O}( \t G)$ for the QED$_2$ type case
\cite{Nandori_LSG,LSG,Nandori_Mschw,Nandori_JPA}.
This is achieved by linearizing the RG equation in the periodic piece 
of the blocked potential, 
\beq
\label{uv_rg}
(2+k\partial_k)\t V_k = -  \frac1{4\pi} \, \,  \frac{F_1(\t V_k)}{C}+\ord{\t V_k^2},
\eeq
where $C$ and $F_1(\tilde U_k)$ stand for the constant and linear 
pieces of the determinant 
\beq
\det[\delta_{ij} + {\tilde V}^{ij}_k] = C + F_1(\t V_k) + {\cal O}(\t V_k^2).
\eeq
The UV scaling law for the QCD$_2$ type LSG model for $N$ colors is
\beq
\label{qcd2yUV}
\t y(k) =  \t y(\Lambda) 
\left(\frac{k}{\Lambda}\right)^{\frac{\beta^2}{N 4 \pi} - 2}
\left(\frac{k^2 + NJ}{\Lambda^2 + NJ}\right)^{\frac{(N-1)\beta^2}{N 8\pi}}
\eeq
with the initial value $\t y(\Lambda)$ at the UV cutoff $k = \Lambda$.
From the extrapolation of the UV scaling law \eq{qcd2yUV} towards the IR scales 
we can read off the critical value $\beta^2_{c}(N) =8\pi N$.
The coupling $\t y$ is irrelevant for $\beta^2>\beta^2_{c}(N)$ and relevant for
$\beta^2<\beta^2_{c}(N)$. The critical frequency and the corresponding critical temperature
\begin{equation}
\label{laydep_j}
T^{(N)}_{\rm{QCD}} = \frac{2\pi}{\beta_c^2 (N)} = 
T^{\star}_{\rm{KTB}} \frac{1}{N}
\end{equation}
separating the two phases of the model coincide with the general expressions
obtained previously for the rotated LSG model in Refs. \cite{Nandori_LSG,Jentschura,Nandori_JPA,Nandori_Mschw}.

Similar consideration can be done for the QED$_2$ type LSG model \cite{Nandori_LSG,LSG,Nandori_Mschw,Nagy_MQED2}
and the solution of \eqn{uv_rg} for the couplings in case of $N$ flavors is given as
\beq\label{qed2yUV}
\t y(k) = \t y(\Lambda) 
\left(\frac{k}{\Lambda}\right)^{\frac{(N-1) \beta^2}{N 4 \pi} - 2}
\left(\frac{k^2 + N G}{\Lambda^2 + N G}\right)^{\frac{\beta^2}{N 8\pi}}.
\eeq
The critical frequency and the corresponding critical temperature which separates the two phases
of the model can be read off directly,
\beq
\label{laydep_m}
\beta^2_{c}(N) = \frac{8\pi N}{N-1},~ \to ~
T^{(N)}_{\rm{QED}} = \frac{2\pi}{\beta_c^2 (N)} =
T^{\star}_{\rm{KTB}} \frac{N-1}{N}.
\eeq
For $N=1$ layer the LSG model with magnetic type coupling reduces to the massive 2D-SG model
where the periodicity is broken explicitly. Therefore, for sufficiently small bare coupling
$\t y(\Lambda)$, there exists only a single phase \cite{Nandori_LSG,Nandori_JPA,LSG,Nandori_Mschw}
in the $N=1$ layer model, i.e., the coupling $\tilde y(k)$ is relevant
(increasing) in the IR limit ($k\to 0$) irrespectively of $\beta^2$.
For $N\to\infty$ the magnetically coupled LSG behaves like a
massless 2D-SG model with the critical frequency $\beta_c^2 = 8\pi$ \cite{Nagy_MQED2}.
Both the dependencies in \eqns{laydep_j}{laydep_m} of the critical frequencies
$\beta_c$ on the number of layers $N$ indicate that QED$_2$ and QCD$_2$ with
any number of flavors and colors, respectively belong to the symmetry broken phase of
the corresponding LSG model ($\beta^2=4\pi<\beta_c^2(N)$), see also \fig{bcrit}.
We shall show below that the IR scaling laws confirm this statement.
\FIGURE{
\epsfig{file=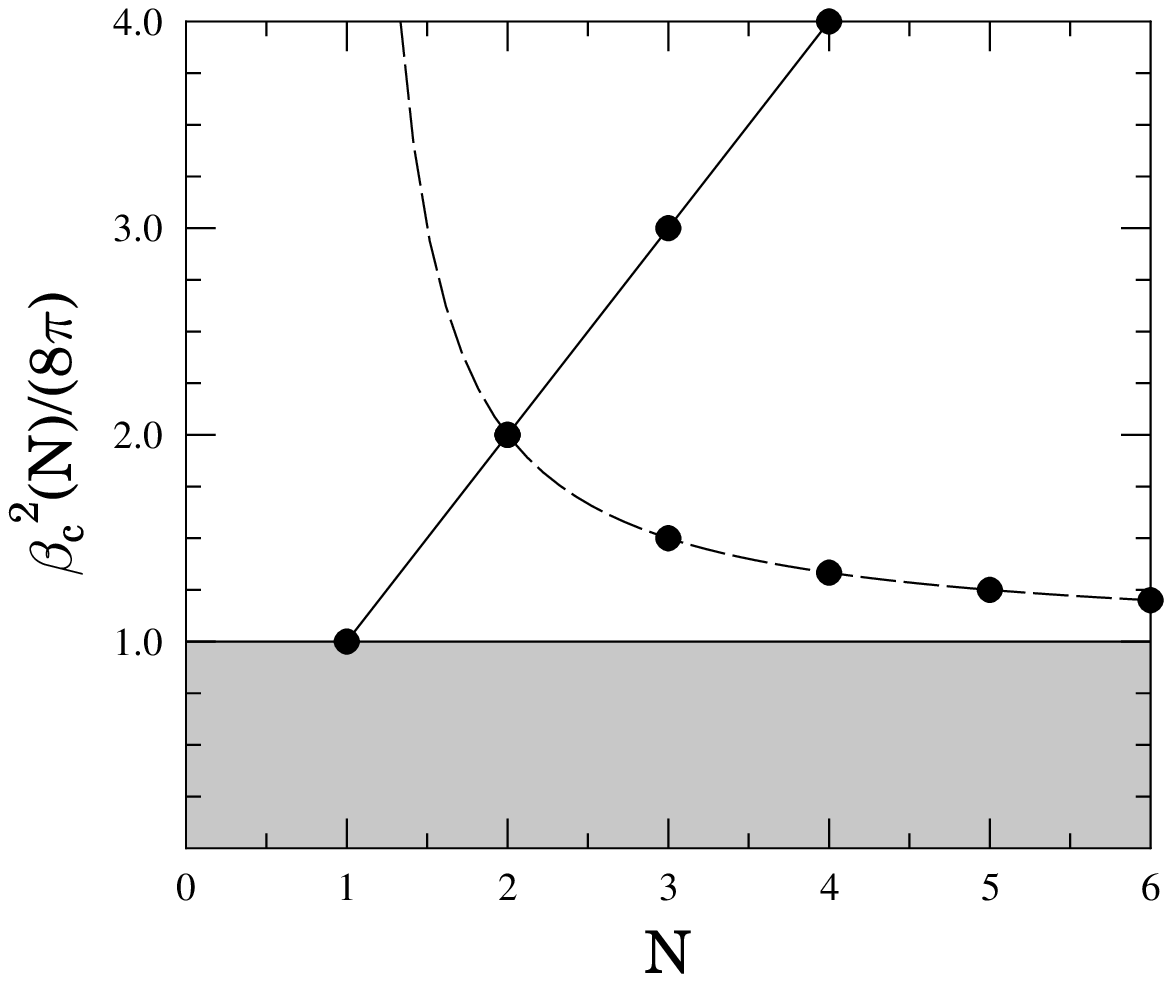,width=9cm} 
\caption{The  critical frequency $\beta_c^2(N)$
versus the  layer-number $N$
is shown for  LSG models with QED$_2$-type (dashed line) and  QCD$_2$-type
(solid line) interlayer couplings, respectively.
 The critical frequencies lie outside of 
the shaded area, irrespectively of $N$. }
\label{bcrit}
}
The UV treatment suggests, that the phase structure of the LSG type models and the SG model is quite
similar, since there is a Coleman fixed point with the critical parameter $\beta_c$ separating the
symmetric $(\beta>\beta_c)$ and the  symmetry broken $(\beta<\beta_c)$ phases. The Coleman point
appears in the UV level, namely the couplings start to scale irrelevantly (relevantly) in the
(broken) symmetric phases respectively. The inclusion of the higher harmonics modifies this picture
by introducing further critical $\beta$ values \cite{Pangon_SG}, and may give further fixed points.
However the IR behavior of the SG model results in a single Coleman point, suggesting the IR nature
of the fixed point.

Occurring a pole  during the RG evolution may give another signal that the bosonized 
multi-flavor QED$_2$ and multi-color QCD$_2$ correspond to
LSG models in the symmetry broken  phase. The following search for such a pole
relies on the extrapolation of the UV scaling laws again.
Nevertheless, the appearance of a pole would inform one on the
superuniversality of the effective potential settling it as \eq{parab} 
with the consequence that the original multi-color QCD$_2$ exhibits
a single phase.

As stated earlier,  QED$_2$ for $N_f=2$ and QCD$_2$ for $N_c=2$ coincide. It was shown
\cite{Nagy_MQED2} that these models have a single phase. The situation changes for
$N_f=3$ and $N_c=3$, respectively. Then the RG equation in \eqn{eveq} for 3 layers has the form
\bea
\label{lsg_3l}
(2+k\partial_k)\t V_k &=&-\frac1{4\pi} \log\biggl[
(1+\t V_k^{11})(1+\t V_k^{22})(1+\t V_k^{33})
+\t V_k^{12} \t V_k^{23} \t V_k^{31}+\t V_k^{13} \t V_k^{21} \t V_k^{32}\nn
&&-\t V_k^{13}(1+ \t V_k^{22}) \t V_k^{31}- \t V_k^{12} \t V_k^{21}(1+ \t V_k^{33})
- (1+\t V_k^{11}) \t V_k^{23} \t V_k^{32}
\biggr].
\eea
For the 3-flavor QED$_2$ the potential for the corresponding scalar model is
\beq
\t V_k = \hf \t G(\phi_1-\phi_2+\phi_3)^2
+\t y\bigl[\cos(\beta\phi_1)+\cos(\beta\phi_2)+\cos(\beta\phi_3)\bigr]
\eeq
with $\t G>0$.
Inserting this ansatz into the argument of the logarithm in the right hand side of the
RG equation \eq{lsg_3l} one can see, that a pole for $\phi_1=\phi_2=\phi_3=0$ may appear when
\beq
(1-\beta^2 \tilde y)^2(1+3\tilde G-\beta^2 \tilde y) = 0.
\eeq
Due to the factor in the first bracket in the left hand side,
the relevant scaling of the dimensionless coupling $\tilde y$ in \eqn{qed2yUV} drives
the flow to a pole independently of the UV initial parameters of the model.
This simple treatment suggests that the model is in its symmetry broken  phase, which implies
that the 3-flavor QED$_2$ has a single phase.
Thorough calculations showed the same result \cite{Nandori_Mschw,Nagy_MQED2}.

The form of the potential for the 3-color QCD$_2$ is
\beq
\t V_k = \hf \t J[(\phi_1-\phi_2)^2+ (\phi_2-\phi_3)^2+ (\phi_3-\phi_1)^2 ]
+\t y [\cos(\beta\phi_1)+\cos(\beta\phi_2)+\cos(\beta\phi_3)]
\eeq
with $\t J>0$.
The argument of the logarithm in \eqn{lsg_3l} is now
\beq\label{3QCD2log}
(1-\beta^2 \tilde y)(1+3\tilde J-\beta^2 \tilde y)^2=0,
\eeq
which also gives a pole due to the first bracket in the left hand side if we consider the 
relevant scaling of the coupling $\t y$ according to \eqn{qcd2yUV}.
Similarly to the result obtained for the 3-flavor QED$_2$ the model seems to be in the
symmetry broken phase giving again a single phase  for the original 3-color QCD$_2$.
Below we shall show that our expectation of a single phase for 3-color
QCD$_2$ is justified by the IR scaling laws.

\subsection{IR scaling}

A more reliable information on the phase structure of the multi-color QCD$_2$ should
be deduced from the IR scaling laws of the corresponding LSG type model.
The proper treatment of the problem requires to find the solution of the rather
complicated partial differential equation \eq{eveq}.
Instead of treating that task in its full complexity, we shall invent the
following strategy. First we perform an $O(N)$ rotation ${\cal R}$ which diagonalizes
the symmetric mass matrix by the rotated field variables $\alpha_i={\cal R}_{ij}\phi_j$.
Due to the particular structure of the mass matrix ${\cal M}^2_{QCD}$, it exhibits a single
zero eigenvalue and $N-1$ identical eigenvalues $N{\t J}$.
Further on, we assume that the mass gap suppresses large amplitude quantum fluctuations
of the massive field components and, therefore   the potential can be Taylor-expanded  in
the  massive field components at their vanishing value. Then the massive fields appear in  
the lowest order as free fields and decouple from the massless field component and can
easily be integrated out. In this manner the problem becomes amenable
for the numerical treatment. Such an approach has been successfully applied to LSG models
in \cite{Jentschura}.

In particular, the rotation for $N=3$  is performed with the matrix
\beq
\label{rot_matrix}
{\cal R} = 
\begin{pmatrix}
\frac{1}{\sqrt{3}} & \frac1{\sqrt{3}} &\frac1{\sqrt{3}}\\
-\frac1{\sqrt{2}} & 0 & \frac1{\sqrt{2}}\\
\frac1{\sqrt{6}} &-\frac{\sqrt{2}}{\sqrt{3}} &\frac1{\sqrt{6}}\\
\end{pmatrix}
\eeq
and the effective action of the rotated model takes the form
\beq
\label{rot_qcd2} 
\Gamma_{\mr{rot}}= \int_x
\left[ \hf (\partial_\mu \alpha_1)^2+\hf (\partial_\mu \alpha_2)^2+ \hf (\partial_\mu \alpha_3)^2+
\frac32 \t J (\alpha_2^2+\alpha_3^2)
+ V_{\mr{rot}}
\right],
\eeq
with
\bea
V_{\mr{rot}} &=& 2 \t y \cos\left(\beta \frac{\alpha_1}{\sqrt{3}}\right)
\cos\left(\beta \frac{\alpha_2}{\sqrt{2}}\right)
\cos\left(\beta \frac{\alpha_3}{\sqrt{6}}\right)
-2 \t y \sin\left(\beta \frac{\alpha_1}{\sqrt{3}}\right)
\cos\left(\beta \frac{\alpha_2}{\sqrt{2}}\right)
\sin\left(\beta \frac{\alpha_3}{\sqrt{6}}\right)\nn
&&+\t y \cos\left(\beta \frac{\alpha_1}{\sqrt{3}}\right)
\cos\left(\beta \frac{\alpha_2}{\sqrt{6}}\right)^2
-\t y \cos\left(\beta \frac{\alpha_1}{\sqrt{3}}\right)
\sin\left(\beta \frac{\alpha_2}{\sqrt{6}}\right)^2\nn
&&+\t y \sin\left(\beta \frac{\alpha_1}{\sqrt{3}}\right)
\cos\left(\beta \frac{\alpha_2}{\sqrt{6}}\right)
\sin\left(\beta \frac{\alpha_2}{\sqrt{6}}\right).
\eea
Now we keep the lowest-order term of the Taylor-expansion of the potential in  the massive
field components at  $\alpha_2=\alpha_3=0$ like in \cite{Jentschura} for the LSG model.
Then  the effective action  reduces to
\beq\label{effgam}
\Gamma^{\mr{red}}_{\mr{rot}} = \int_x\left[\hf (\partial_\mu\alpha_1)^2
+3 \t y \cos\left(\beta \frac{\alpha_1}{\sqrt{3}}\right)\right].
\eeq
The massive components decouple and describe free fields, and by integrating out the $N-1$ 
massive modes as described in Refs.~\cite{Jentschura,Nandori_Mschw} one obtains that the IR
behavior of the model is completely determined by the remaining massless component field
$\alpha_1$ which describes a simple SG model.

The advantage of the rotation reveals itself in the reduction of
the task of the low-energy QCD$_2$ to the determination of the IR behavior of
the SG model \eq{effgam}. The functional RG method showed \cite{Nandori,Nagy_SG,Nagy_ZSG}
that this SG model has two phases depending on the value of its parameter
$\beta^2/3$. As we have shown previously in Sect. \ref{uvscal} , the multi-color QCD$_2$ 
gives values of the parameter $\beta^2=4\pi<\beta_c^2(N)$ for arbitrary number $N>1$ of colors,
which  means that the corresponding SG model is in the symmetry broken  phase.
Earlier calculations based on Fourier expansion \cite{Nandori,Nagy_JPA,Nagy_SG,Nandori_scheme} showed that
in the symmetry broken  phase the effective potential is superuniversal with the parabolic
shape \eq{parab}. It happened numerically that the Fourier-expansion drove the evolution
towards the pole at a non-vanishing scale. Approaching it a parabolic prepotential appeared
but the Fourier-expansion became unreliable at the same time, so that 
the further evolution was treated at tree level \cite{Polonyi_tree}
which always gave parabolic effective potential \eq{parab}.
More precise calculations, avoiding any expansion of the potential \cite{Pangon_SG} now suggest,
however,  that there is a non-trivial IR attractive fixed
point for low values of $\beta^2$ giving a superuniversal effective potential which 
deviates  a little from \eqn{parab}, and the latter form is  reached only in the limit $\beta^2\to 0$.

The flow of the coupling $\t y$ is determined by a computer algebraic program \cite{Nandori_multiSG},
which solves the RG equation directly, without using any ansatz for the potential.
It finds the fundamental mode $\t y(k)$ by Fourier-analyzing the numerically determined potential at any
scale $k$ afterwards. The polynomial suppression scheme we use here needs higher numerical
accuracy as compared to the exponential scheme \cite{Berges}. We set a high numerical working precision
in order to handle the numerical ambiguities properly as was pointed in \cite{Pangon_SG}.
The IR flow of the coupling can be seen in \fig{fig:yk}.
\FIGURE{
\epsfig{file=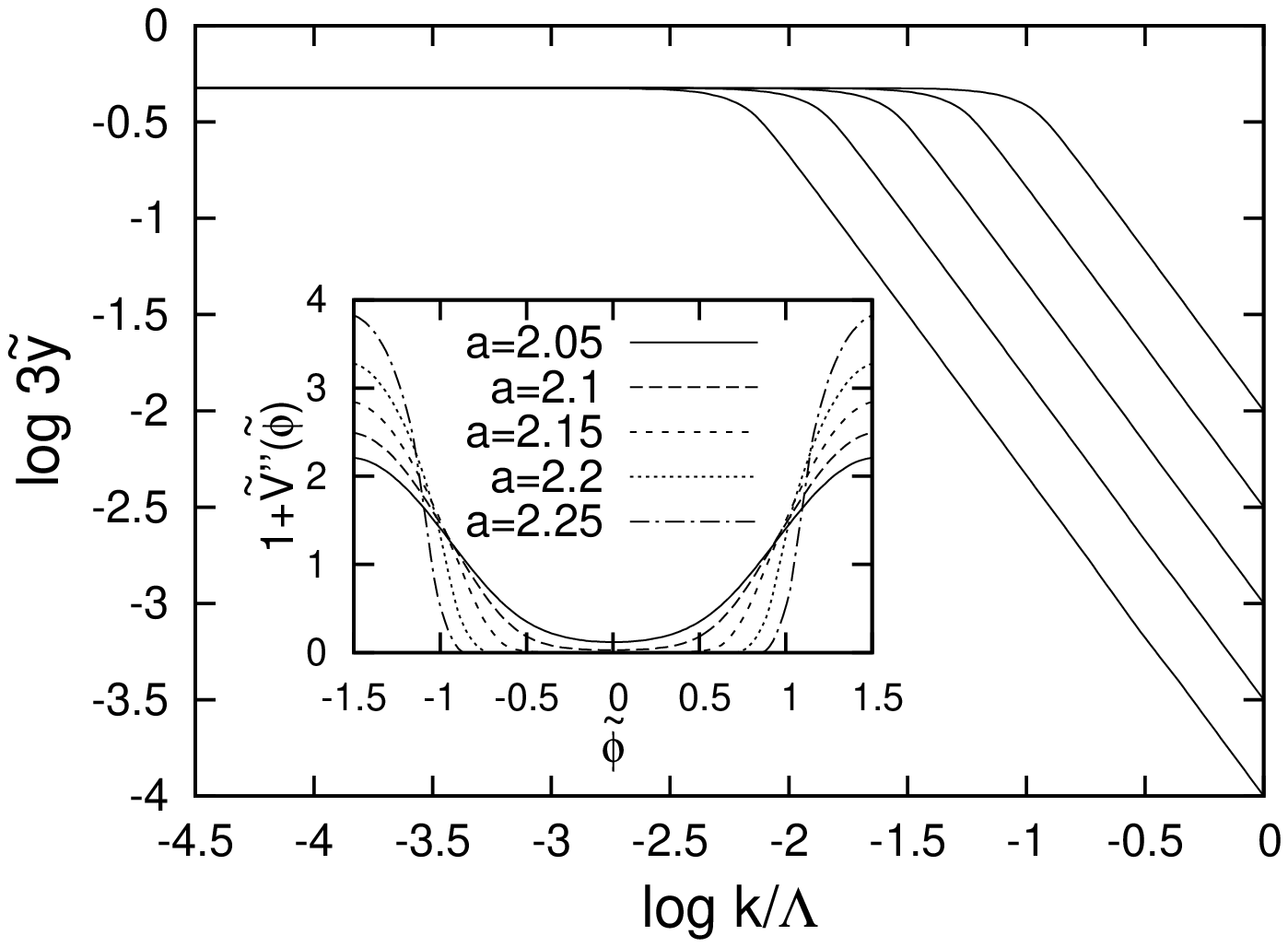,angle=-90,width=10cm} 
\caption{The IR scaling of the dimensionless coupling $\t y$ for several initial values.
The inset demonstrates the convexity of the effective action during the evolution as
the scale $k$ decreases, $k/\Lambda=10^{-a}$.}
\label{fig:yk}
}
The figure clearly shows that the IR value of the dimensionless coupling $\t y(0)$ is independent
of its initial, microscopic value. This implies that the IR effective potential is also
independent of the microscopic parameters, i.e. it is superuniversal. According to the
inset of \fig{fig:yk} the effective potential approaches the parabolic shape characterized
by $1+\t V''(\Phi)=0$ in the limit $k\to 0$, but our high-precision calculation also shows
that it does not reach the parabola, in accordance with the findings in \cite{Pangon_SG}.

The dimensionful parameter $J=g^2c_g$ is constant, therefore the fermionic coupling
$g$ remains unchanged during the evolution. It gives non-vanishing coupling
in the IR limit of the 3-color QCD$_2$. The IR behavior of the model shows that the
dimensionful coupling $y$ goes to zero for $k\to 0$ driving the quark mass $m_g$ to zero.
Therefore this scalar model is a free massive theory too as was the 3-flavor QED$_2$ \cite{Nagy_MQED2},
and the 3-color QCD$_2$ is an interacting theory of massless two-dimensional quarks.

\subsection{Large $N$ case}

The techniques used in the previous subsections can be easily generalized to the case of
arbitrary $N$. The naive expectation that multi-color QCD$_2$ is in the symmetry broken phase stems from
the appearance of poles  if we consider the argument of the logarithm in the right hand side in \eqn{3QCD2log}
for $N$ colors,
\beq\label{npole}
(1-\beta^2 \tilde y)(1+N\tilde J-\beta^2 \tilde y)^{N-1}=0
\eeq
in the framework of the extrapolation of the UV scaling laws.
Similarly to the 3-color case, the first factor can change sign due to the relevant UV scaling of the coupling
$\t y$, signaling the appearance of the pole and the symmetry broken  phase. However, a more reliable
conclusion can be drawn again if one considers the IR scaling  for $N$-colors. In order to diagonalize
the mass matrix, one performs the appropriate $O(N)$ rotation. Taylor-expanding the potential in the
new massive field variables $\alpha_i$, $i=2\ldots N$ at their vanishing values and keeping the quadratic
terms only, these become massive free fields. Integrating them out one can reduce the effective action to
that of the single massless  field $\alpha_1$,
\beq\label{effgamN}
\Gamma^{\mr{red}}_{\mr{rot}} = \int_x\left[\hf (\partial_\mu\alpha_1)^2
+N \t y \cos\left(\beta \frac{\alpha_1}{\sqrt{N}}\right)\right].
\eeq
This  is a SG model too with decreasing parameter $\beta'=\beta/\sqrt{N}$ for increasing number $N$ of colors.
The calculation for smaller values of $\beta'$ requires extreme accuracy. We set the working precision
to several hundreds during the calculations to get some reliable numerical information for the model.
If the effective potential is a parabola in \eqn{parab} then the IR value of the dimensionless coupling is
\beq\label{yir}
\t y = \frac2{\beta^2}.
\eeq
If we take the parameters from \eqn{effgamN}, then this relation becomes the same. 
Let us note here, that the IR value \eq{yir} differs a factor of 2 of the one for which the
pole occurs in \eqn{npole}. This is due to the circumstance that the relation \eq{npole}
corresponds to the neglection of the higher-harmonics of the blocked periodic potential,
whereas our numerical approach avoiding the Fourier-expansion takes automatically all higher harmonics with.
In order to enlighten the deviation of our result from the Maxwell-cut induced parabolic effective
potential \eq{parab} it is reasonable to plot $2-\t y\beta^2$ as the function of the color $N$, see \fig{fig:nc}.
We succeeded to calculate it only up to $N=5$. Nevertheless the figure clearly shows that
$\t y\to 2/\beta^2$ for $N\to\infty$.
\FIGURE{
\epsfig{file=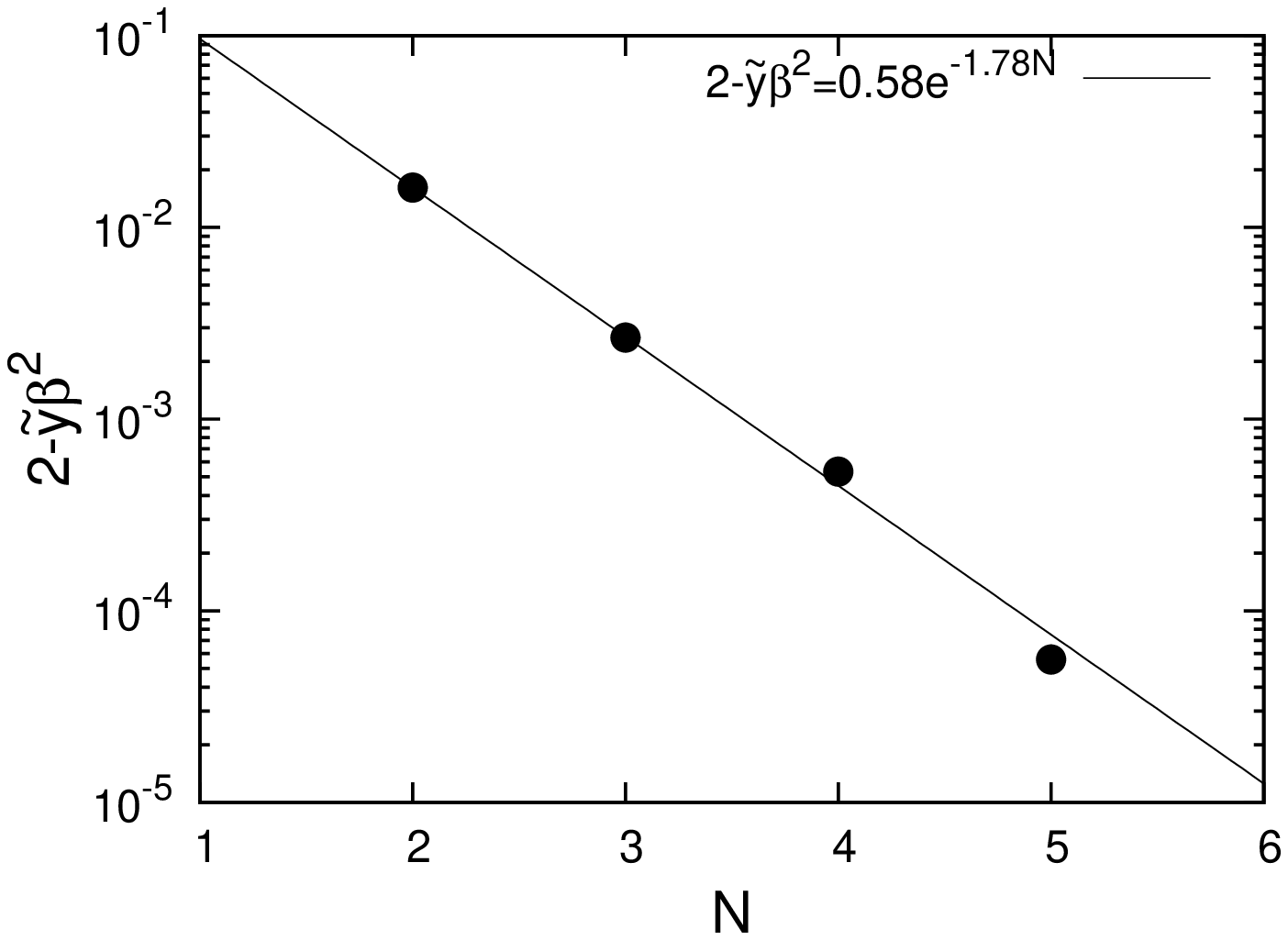,angle=-90,width=10cm} 
\caption{The $N$ dependence of the function $2-\t y\beta^2$.}
\label{fig:nc}
}
These results also give vanishing dimensionful coupling $y$ in the IR limit. One can have
the same conclusion as was obtained in the 3-color case, namely the low energy $N$-color QCD$_2$ is
a massless interacting theory.

\section{Summary}\label{sec:sum}

The phase structure of the bosonized version of the multi-color QCD$_2$ is mapped out
and it was shown, that the model possesses a single phase.

After bosonization of the original fermionic model a periodic self-interacting
scalar model is obtained with a mass matrix. The more involved last self-interaction term 
of the Hamiltonian \eq{qcdbos} is shown to be treatable by Taylor-expansion giving further corrections to the mass
matrix, so far the system is considered in or close to the ground state.
The bosonized 2-color QCD$_2$ coincides with the 2-flavor QED$_2$
which implies the similar trivial phase structures of these models. The bosonized
3-color QCD$_2$ and 3-flavor QED$_2$ are, however, different models. The bosonized
3-flavor QED$_2$ is known to be in the symmetry (periodicity) broken phase and represents a free massive theory.
The scaling laws and the phase structure of the bosonized multi-color QCD$_2$ has been determined 
by the functional RG technique, applying the Callan-Symanzik renormalization scheme.
The periodic self-interaction has been found to be UV relevant, which tries to drive the flow into
a pole, signaling that the $N$-color QCD$_2$ is in the symmetry broken phase.
The IR physics of the $N$-color QCD$_2$ has been also determined after
diagonalizing the mass matrix and integrating out the massive fields in the free-field approximation.
It was found that the bosonized multi-color QCD$_2$ represents effectively a SG-type model
in the symmetry broken phase, characterized by a superuniversal dimensionless effective potential.
In the IR limit the quantum fluctuations  try to drive the system to a non-trivial saddle point
what seems to be, however, never reached at finite energy scale. Nevertheless, the larger the
number $N$ of colors  is the closer the dimensionless effective potential is driven to the parabolic
shape \eq{parab}. Making use of  the bosonization relations, we have concluded that the $N$-color
QCD$_2$ is a massless interacting fermionic model.

\section*{Acknowledgements}

Our work is supported by T\'AMOP 4.2.1-08/1-2008-003 project.
The project is implemented through the New Hungary Development Plan co-financed
by the European Social Fund, and the European Regional  
Development Fund.

\end{document}